\documentclass[fleqn,twoside]{article}
\usepackage{espcrc2}
\usepackage{graphicx}
\usepackage[figuresright]{rotating} 
\usepackage[square,comma,numbers]{natbib} 

\newcommand{\dal}{\ensuremath{\Delta \alpha/ \alpha}}

\def\kms {\hbox{${\rm km\ s}^{-1}$}}
\def\scm  {$\hbox{{\rm cm}}^{-2}$}    
\def\MOLH {\hbox{${\rm H}_2$}}  
\def \HI {H{\sc \,i}}

\def\lapp{\ifmmode\stackrel{<}{_{\sim}}\else$\stackrel{<}{_{\sim}}$\fi}
\def\gapp{\ifmmode\stackrel{>}{_{\sim}}\else$\stackrel{>}{_{\sim}}$\fi}

 \newcommand{\AmS}{{\protect\the\textfont2
  A\kern-.1667em\lower.5ex\hbox{M}\kern-.125emS}}

\title{Measuring Changes in the Fundamental Constants with Redshifted Radio Absorption Lines}

\author{S. J. Curran\address[UNSW]{School of Physics, University of New South Wales, Sydney, Australia},
N. Kanekar\address{Kapteyn Institute, University of  Groningen, The Netherlands}, J. K. Darling\address{Carnegie Observatories, Pasadena, USA}}

\begin{document}

\begin{abstract}
Strong evidence has recently emerged for a variation in the fine
structure constant, $\alpha\equiv e^2/\hbar c$, over the history of
the Universe.  This was concluded from a detailed study of the
relative positions of redshifted optical quasar absorption
spectra. However, {\it radio} absorption lines at high redshift offer
a much higher sensitivity to a cosmological change in $\alpha$ than optical
lines. Furthermore, through the comparison of various radio
transitions, \HI, OH and millimetre molecular (e.g. CO) lines, any
variations in the proton g-factor, $g_p$, and the ratio of
electron/proton masses, $\mu\equiv m_e/m_p$, may also be
constrained. Presently, however, systems exhibiting redshifted radio
lines are rare with the bias being towards those associated with
optically selected QSOs. With its unprecedented sensitivity, large
bandwidth and wide field of view, the SKA will prove paramount in
surveying the sky for absorbers unbiased by dust extinction. This is
expected to yield whole new samples of \HI ~and OH rich systems, the
latter of which will prove a useful diagnostic in finding redshifted
millimetre absorbers. As well as uncovering many new systems through
these blind surveys, the SKA will enable the detection of \HI
~absorption in many more of the present optical sample -- down to
column densities of $\sim10^{17}$ \scm, or $\gapp2$ orders of
magnitude the sensitivity of the current limits. Armed with these
large samples together with the high spectral resolutions, available
from the purely radio comparisons, the SKA will provide statistically
sound measurements of the values of these fundamental constants in the
early Universe, thus providing a physical test of Grand Unified
Theories.

\end{abstract}

\maketitle

\section{INTRODUCTION}
\label{intro}

Much interest has recently centered on the possibility that
fundamental constants such as the fine structure constant might vary
with cosmic time. Many theoretical models, such as superstring
theories (presently the best candidate to unify gravity and the other
fundamental interactions) and Kaluza-Klein theories, {\it naturally
predict a spatio-temporal variation of these constants}. Such models
usually invoke extra dimensions compactified on tiny scales, with the
fundamental constants arising as functions of the scale lengths of
these extra dimensions (e.g.  \cite{mar84,dp94}).  At present, no
mechanism has been found to keep the compactified scale-lengths
constant implying that, if extra dimensions exist and their sizes
undergo cosmological evolution, the three dimensional coupling constants
should vary with time (e.g.  \cite{lg98}). Several other modern
theories also provide strong motivation for an experimental search for
variation in the fine structure constant (e.g.
\cite{ww86,hr88}). Interestingly, varying constants provide
alternative solutions to cosmological fine-tuning problems, such as
the ``flatness problem'' and the ``horizon problem'' \cite{sbm02}.

From terrestrial experiments, \cite{fif+00} find $\dal < 1.2\times
10^{-7}$, whereas $\geq4.5\times 10^{-8}$ is found by \cite{lam03},
each from isotopic abundances measured in the Oklo natural fission
reactor. This is $1.8 \times 10^9$ years old, corresponding to a
redshift of $z\sim0.1$.

However, astrophysical studies of redshifted spectral lines provide a
powerful probe of putative changes in fundamental constants over a
large fraction of the age of the Universe
(e.g.~\cite{wfc+98,cms+00,iprv03,scpa04}):  Recent studies of the
relative redshifts of metal-ion atomic resonance transitions in the
optical (Keck/HIRES) quasar spectra of 143 heavy element absorption
systems are consistent with a smaller fine structure constant in the
intervening absorption clouds over the redshift range $0.2 < z_{\rm
abs} < 3.7$\begin{figure}[h]
\vspace{4.9 cm} \setlength{\unitlength}{1in} 
\begin{picture}(0,0)
\put(-0.15,-1.85){\includegraphics{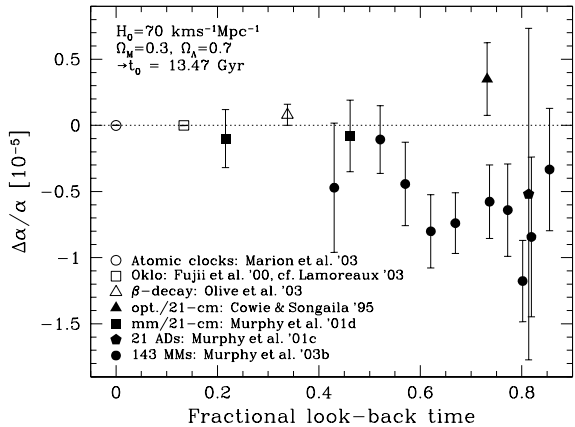}}
\end{picture}
\caption{Summary of the previous  measurements of $\Delta\alpha/\alpha$.
The open symbols are local constraints \protect\cite{fif+00,lam03,mpa+03,opy+03} and the filled
symbols are QSO absorption constraints from the alkali doublet \protect\cite{mwf+01c},
the many-multiplet \protect\cite{mwf03} and the radio methods (discussed
in this chapter). Courtesy of Michael Murphy.}
\label{f1}
\end{figure}
(Fig.\ref{f1}). However, since all of the observations which suggest a
change in $\alpha$ use the same telescope and instrument \cite{mwf03},
despite our best efforts \cite{mwf+01b}, unforeseen systematic effects
cannot be ruled out. Furthermore, \cite{cspa04} find no such variation
from VLT/UVES data of 23 systems ($0.4 < z_{\rm abs} < 2.3$). Given
the implications for theoretical physics, it is clear that independent
checks are required which can refute or confirm the optical
results. In particular, radio studies offer:
\begin{enumerate}
  \item A different technique of potentially much higher accuracy than
	the purely optical methods.
    \item The ability to study the evolution of other dimensionless
	fundamental constants, besides $\alpha$.
	\end{enumerate}
Currently, however, the paucity of systems exhibiting H{\sc i} 21 cm
together with rotational or optical absorption severely limits our
ability to carry out statistically sound comparisons. In this chapter
we shall discuss how the SKA is expected to significantly increase the
number of high redshift radio absorbers from which detailed study can
greatly improve constraints in the cosmological variation in the fine
structure constant, the ratio of electron--proton mass, and the proton
g-factor.

\section{RADIO LINES \& FUNDAMENTAL CONSTANTS}
\label{radio}

Spectral lines such as the \HI~21~cm line, cm-wave OH lines and
mm-wave CO and HCO$^+$ lines arise from different physical mechanisms
and hence have different dependences on these constants
(e.g. \cite{dar03,ck03,mwf+00,kc04}). Further, the spectral resolution
commonly available in radio spectroscopy is typically far better than
that obtained in the optical domain, allowing higher precision
redshift measurements.

Consider two transitions, with rest
frequencies $\nu_1(z)$ and $\nu_2(z)$ that depend on redshift through
their (different) dependences on certain fundamental constants. The
ratio of the two frequencies, $y \left( \alpha, g_p, \mu,... \right)
\equiv \nu_1/\nu_2$ will vary with redshift as:
\begin{equation}
\label{eqn:basic}
\frac{\Delta y}{y}  = \frac{y\left(z\right) - y_0}{y_0} \approx 
\frac{\Delta z}{1 + z_1} = \frac{z_2 - z_1 }{1 + z_1},
\end{equation}
where $z_1$ and $z_2$ are the {\it measured} redshifts of the two
lines in a given object and $y_0$ is the value of the ratio at the
present epoch. One could thus, in principle, compare the redshift of
an object as measured from the \HI~21~cm, OH, and millimetre
transitions, as well as the optical fine structure lines, in order to
determine the evolution of combinations of different
constants. Clearly, this must be carried out on a statistically large
sample of objects in order to average out any intrinsic velocity
offsets between the different species or differing lines of sight
between the mm-wave and cm-wave continua.

\subsection{Atomic Lines}
\label{HI}

The \HI~21 cm line arises from a hyperfine spin-flip transition
($\nu_{21} = 1420.406$ MHz), produced by the interaction of the
magnetic fields of the electron and proton. The line frequency has the
dependence
\begin{equation}
\label{eqn:21cm}
\nu_{21} \propto g_p \mu \alpha^2 R_\infty\:,
\end{equation}
where $R_\infty \equiv m_e e^4/\hbar ^3 c$ is the Rydberg 
constant\footnote{Note 
that the frequencies of all transitions discussed here are proportional 
to $R_\infty$. Since this cancels out when comparing frequency ratios, 
it is usually not quoted in the literature.}.
By comparison, the frequency of the optical
transitions, governed by Coulombic interactions, are
\begin{equation}
\nu_{\rm opt}\propto (1 + 0.03\alpha^2)R_\infty,
\end{equation}
thus we see that the spin-flip transition is $\approx30$ times as
sensitive than the optical resonance transition to a given change in
$\alpha$ \cite{dfw98b}. The ratio of frequencies is
\begin{equation}
\frac{\nu_{21}}{\nu_{\rm opt}}\propto\mu\alpha^2 g_{p},
\end{equation}
(e.g. \cite{cs95}), and so increasing the number of cosmologically distant 
QSO absorption systems with both optical and 21 cm lines would result in 
a significant improvement in measuring any cosmological changes in $\mu\alpha^2 g_{p}$.

\subsection{Molecular Lines}

\subsubsection{Centimetre-wave OH lines}
\label{OH}

Cm-wave OH lines allow the interesting possibility of simultaneously
measuring changes in $\alpha$, $g_p$ and $\mu$, using spectral lines
of a {\it single species} \cite{dar03,ck03,kc04}. This is because
these lines arise from two entirely different physical mechanisms,
Lambda-doubling and hyperfine splitting \cite{ts55}, which have
different dependences on the above constants: Each OH state gives rise
to four cm-wave lines, allowing the possibility of multiple,
independent estimates of any cosmological evolution. The strongest
transitions are found in the ground state, at frequencies of $\nu_{\rm
rest} =1665.402$~MHz and 1667.359~MHz (``main'' lines) and
1612.231~MHz and 1720.530~MHz (``satellite'' lines). At present, these
are the only Lambda-doubled lines to have been detected at
cosmological distances \cite{cdn99,kc02a} and hence, the best
candidates for searches at higher redshifts with the SKA. It should be
emphasized, however, that the SKA will cover the frequencies of {\it
all} known Lambda-doubled OH lines (i.e.  rest frequencies of 1.6 to
13 GHz) with high sensitivity.

An interesting aspect of the OH lines is the fact that various line frequency
differences depend on all three of the above constants, while the
frequency sums only depend on $\alpha$ and $\mu$. For example, in the
case of the ground state 18 cm lines, we have \cite{ck03}
\begin{eqnarray}
\label{eqn:OH18cm1}
\nu_{1665} + \nu_{1667} &\propto& \mu^{2.57} \alpha^{-1.14},\\
\label{eqn:OH18cm2}
\nu_{1667}-\nu_{1665}  &\propto& \mu^{2.44} \alpha^{-0.88} g_p, \\
\label{eqn:OH18cm3}
\nu_{1720}-\nu_{1612} &\propto& \mu^{0.72} \alpha^{2.56} g_p, 
\end{eqnarray} 
where we see that Equations (\ref{eqn:OH18cm1}) and (\ref{eqn:OH18cm3}) have
very different dependences on both $\alpha$ and $\mu$, giving the
ratio of the two frequencies a high sensitivity to changes in their
numerical values.

In the next 
excited state, $^2\Pi_{1/2}\ J = 1/2$, only the 4750.656~MHz main line 
is allowed (the $F = 0\rightarrow0$ main line is forbidden).
This has the dependence \cite{kc04}
\begin{equation}
\label{eqn:OH6cm}
\nu_{4750} \propto  \mu^{0.51} \alpha^{2.98} R_\infty \:\: .
\end{equation}
Again, Equations~(\ref{eqn:OH18cm1}) and (\ref{eqn:OH6cm}) have very 
different dependences on $\alpha$ and $\mu$: In fact, their ratio is 
proportional to $\alpha^{4.12}$, one of the most sensitive to a change in 
the fine structure constant.

\subsubsection{Millimetre-wave molecular lines}
\label{CO}

Mm-wave molecular lines from species such as CO, HCO$^+$, etc. arise
from rotational transitions (e.g. \cite{wc97}), with line frequencies
proportional to $\mu R_\infty$, thus providing a useful ``anchor''
with which to compare the \HI ~and OH transitions. The high spectral
resolution available makes millimetre lines potentially very useful and
they may be compared with the 21 cm line giving
\cite{dwbf98}
\begin{equation}
\frac{\nu_{21}}{\nu_{\rm mm}}\propto\alpha^2 g_{p},
\end{equation}
or with the OH lines, giving\\
\begin{equation}
\frac{\nu_{1667}}{\nu_{\rm mm}}\propto\mu^{1.57}\alpha^{-1.14}\ ,\ 
\frac{\nu_{4750}}{\nu_{\rm mm}}\propto\mu^{-0.49}\alpha^{2.98}.
\end{equation}
Regarding these last constraints, a fairly tight linear correlation has
been found between HCO$^+$ and OH column densities, both in the Galaxy
\cite{ll96a} and out to $z \sim 1$ \cite{kc02a}, extending over more
than two orders of magnitude in column density. This suggests that
HCO$^+$ and OH are likely to be located in the same region of a
molecular cloud, reducing the likelihood of velocity offsets between
the two species. Similarly, \cite{dwbf98} demonstrated a correlation
between the individual velocity components of 21~cm and HCO$^+$
absorption along 9 lines of sight in the Galaxy, with a dispersion of
1.2~\kms.

\subsection{Present radio results}
\label{currentmol}

Only four redshifted radio molecular absorbers are presently known:
$z_{\rm abs} = 0.247$ toward PKS~1413+135; $z_{\rm abs} = 0.674$
toward B3~1504+377; $z_{\rm abs} = 0.685$ toward B0218+357; and
$z_{\rm abs} = 0.889$ toward PKS~1830--210.  The \HI~21~cm
\cite{cps92,cry93,cmrr97,cdn99}, millimetre (e.g. CO)
\cite{wc97,wc95,wc96b,wc98} and 18~cm OH main lines \cite{cdn99,kc02a}
have been detected in all these systems while the 18~cm OH satellite
lines have only been detected towards PKS~1413+135 \cite{kc04}. These
satellite lines have a clear velocity offset ($\approx
12$~\kms~) from the main OH, \HI~and mm-wave lines. Further, the
current \HI~and OH spectra toward PKS~1830--210 and B3~1504+377 have
fairly poor velocity resolutions ($\gapp 10$~\kms~
\cite{kc02a,cdn99,cmr+98}), giving low accuracy measurements. As a
result, constraints on changes in different combinations of
fundamental constants are only available from the lines of sight
toward PKS~1413+135 and B0218+357:
\begin{enumerate}
	\item By comparing VLBA \HI~21~cm redshifts
with those obtained from the CO line, \cite{cms+00}
obtained $\Delta x/x = (1 \pm 0.3) \times 10^{-5}$ at $z_{\rm abs} = 0.685$
towards B0218+357 and $\Delta x/x = (1.3 \pm 0.08) \times 10^{-5}$, where $x \equiv g_p\alpha^2$, at $z\sim 0.247$ towards PKS~1413+135. 
	\item The redshift of the sum of OH~1665 and
1667~MHz lines toward B0218+357 is in agreement with both the
\HI~21~cm and HCO$^+$ redshifts \cite{ck03} and gives the constraint
$\Delta x_1 /x_1 = (-3.5 \pm 4.4) \times 10^{-6}$ (comparing \HI~and
OH) and $\Delta x_2 /x_2 = (6.5 \pm 3.3) \times 10^{-6}$ (comparing
HCO$^+$ and OH), where $x_1 = \mu^{1.571} \alpha^{-3.141} g{^{-1}_p}$
and $x_2 = \mu^{1.571} \alpha^{-1.141}$.  

Note that only statistical errors are included in these estimates:
\cite{cms+00} estimate further systematic errors of $\approx 2 \times
10^{-5}$ towards both B0218+357 and PKS~1413+135, arising from
small-scale motions of $\sim 10$~\kms~.
       \item \cite{mwf+00} carried
out a simultaneous fitting of the single dish \HI~and HCO$^+$ profiles
to obtain $\Delta x/x = (-0.20 \pm 0.44) \times 10^{-5}$ towards
PKS~1413+135 and $\Delta x/x = (-0.16 \pm 0.54) \times 10^{-5}$
towards B0218+357, where $x \equiv g_p\alpha^2$.
	\item Finally, \cite{kcg04} have used the conjugate nature of the 1720 and
1612~MHz satellite lines toward PKS~1413+135 to argue that they arise
in the same gas and obtained $\Delta{x}/x = (2.2 \pm 3.8) \times
10^{-5}$, where $x \equiv g_p [ \alpha^2/y ]^{1.849}$.
\end{enumerate}
In addition to these purely radio results, the best current constraint
from 21~cm and optical comparisons is from the $z_{\rm abs} = 1.77$
absorber towards QSO~1331+170, which gives $\Delta x/x = 0.7 \pm 1.1
\times 10^{-5}$ ($\Delta x/x \equiv \alpha^2 g_p \mu$,
\cite{cs95}\footnote{Again, the error is statistical assuming that no
velocity offset exists between the optical and radio absorbers.}). 

In summary, the current radio studies exhibit no fractional changes in
the fundamental constants of $\gapp 10^{-5}$ to $z_{\rm abs} \sim
0.7$, which is not inconsistent with the optical results.

\section{PROSPECTS WITH THE SKA}
\label{ska}

\subsection{Technical Considerations}
\label{tech}

The primary reason for the relatively few estimates (or constraints)
of changes in different fundamental constants in the radio regime has
been the small number of systems in which radio absorption has been so
far detected (\S \ref{currentmol}). As previously noted, radio
molecular absorption (CO, HCO$^+$, OH, etc) has so far only been
detected in four cosmologically distant absorbers, with the highest
redshift absorber at $z_{\rm abs} \sim 0.9$ \cite{wc98,cdn99}.
Although the situation is somewhat better for 21~cm line absorption
(see \S \ref{targets}), only a handful of absorbers are known at
intermediate redshifts, with just two confirmed detections at $z_{\rm
abs} > 2.1$ \cite{cmr+98,ubc91}.

There are three main reasons for the small size of the radio absorber
sample; low sensitivity, limited frequency coverage and severe radio frequency
interference (RFI). While the former two factors affect observations at all
radio bands, the latter is especially detrimental to observations of
\HI~21~cm and OH 18~cm lines which are redshifted to $\lapp 1$~GHz,
a range strongly contaminated by terrestrial interference. Regarding
the SKA:
	\begin{enumerate}
\item  Sensitivity: The \HI~21~cm and OH transitions are far weaker than the optical 
resonance lines. The high sensitivity of the SKA is thus a critical factor in 
the detection of new redshifted absorption systems. This is also crucial for 
the precise measurements of the absorption redshifts of the different transitions.
The current specifications for the sensitivity are
	\begin{enumerate}
	\item $A_{\rm eff}/T_{\rm sys}= 5000$ at 200~MHz, giving an r.m.s. noise of $\sigma=0.2$~mJy,
	\item $A_{\rm eff}/T_{\rm sys}= 20000$ at 0.5 to 5~GHz, giving $\sigma=0.04$~mJy,
	\item $A_{\rm eff}/T_{\rm sys}= 10000$ at 25~GHz, giving $\sigma=0.01$~mJy, 
	\end{enumerate}
per $1$~\kms~ channel after one hour of integration. By comparison, the most sensitive current low frequency interferometer, the 
GMRT, would have to integrate for $\approx650$~hours at P-band to reach 
similar noise levels.
\item Frequency and sky coverage: The SKA will provide uniform frequency coverage 
from $\approx 100$~MHz to $\approx 25$~GHz \cite{jon04}. That is, it 
will be sensitive to \HI~21~cm absorption and ground state 18~cm OH absorption 
at redshifts of  $z \lapp 13$ and $z \lapp 16$, 
respectively. This coverage also gives HCO$^+$ and CO $J=0\rightarrow1$
absorption at redshifts $z \gapp 2.6$ and $z \gapp 3.6$, respectively. 

The wide-band correlator will provide $10^4$~channels over a bandwidth
25\% of the band centre frequency \cite{jon04}. In combination with
its unrivaled low frequency coverage and the very wide field of view
(200~square degrees at 0.7~GHz \cite{jon04}), this will enable highly
efficient blind surveys for \HI ~and OH absorption, never before
possible with an interferometer.
\item RFI: Terrestrial interference has long been the bugbear of
searches for spectral lines at low frequencies, which lie outside the
protected radio bands. This is particularly severe for systems at high
redshift, for which the \HI~and OH ground state lines lie below 1~GHz,
where TV and radio stations, mobile phones, navigation systems, etc.
make these frequency ranges a veritable ``RFI-forest''. With its
remote location and modern RFI mitigation techniques (such as beam
nulling), the SKA is expected to reach theoretical noise levels in
unprotected radio bands.

\item Resolution: The SKA will have an angular resolution of $\leq
  0.02/\nu_{\rm ~GHz}$ arc-sec \cite{jon04}. This will allow one to
  spatially map the \HI~and OH absorption, ensuring that comparisons
  between different lines are made for similar lines of
  sight. Furthermore, lines of sight towards different source
  components will provide independent measurements of changes in the
  three constants, $\alpha, g_p$ and $\mu$.

\end{enumerate}

\subsection{Possible Targets}
\label{targets}

Of all the known radio transitions, 21 cm is, not surprisingly, the most 
commonly detected at high redshift (cf. \S \ref{currentmol}).
Known redshifted 21~cm absorbers can be broadly separated into three classes: 
\begin{enumerate}
    \item Damped Lyman-alpha absorption systems (DLAs) due to intervening
     galaxies (\cite{kc02} and references therein).  
	\item \HI~clouds associated with the QSO host (e.g. \cite{cps92,cmr+98,kac02,ida03,vpt+03}).
     \item Gravitational lenses (e.g. \cite{mcm98,cdn99,kb03}).
	\end{enumerate}
All of which have exceedingly high neutral hydrogen column densities ($N_{\rm HI} \gapp
10^{20}$~cm$^{-2}$).

Regarding the first class, at present, poor frequency coverage and RFI
have restricted searches for 21~cm absorption to 31 DLAs
\cite{kc02}\footnote{See \cite{cmp+03} for further details.}, with
detections in 16 systems.  Unfortunately, most of these are at low
redshift, with only three detected at $z \sim 2$, all of which with
relatively low quality 21~cm spectra
\cite{wd79,wbj81,wbt+85}. Furthermore, many of the known DLAs were
selected from optical spectra, since these give a broad spectral
coverage and hence enable a large redshift range to be searched
efficiently. Therefore present DLA samples suffer from dust bias,
making these poor candidates in which to search for
molecules\footnote{Also potentially affecting both chemical evolution
and number density evolution studies, although radio selected quasars
and follow-up optical spectroscopy have been used to partly overcome
this problem \cite{eyh+02}.}. The high sensitivity and broad
spectral coverage of the SKA will enable surveys to be carried out entirely in
the radio regime, avoiding all previous dust extinction bias problems
(see Kanekar \& Briggs, this volume). This will be of particular
interest at $z_{\rm abs} \lapp 1.8$, redshifts where the
Lyman-$\alpha$ line cannot be observed by ground-based
telescopes\footnote{Note that current DLA samples contain hardly any
systems in the redshift range $0.7 < z_{\rm abs} < 1.8$
\cite{pgw03a}.}: The cold phase of \HI~is also expected to be more
prevalent at lower redshift due to increased star formation, resulting
in higher metallicities and thus a larger number of radiation pathways
for the gas to cool \cite{kc02}.

Many of the latter two classes are often\footnote{Although not
always. For example, \cite{vpt+03} find 21 cm absorption in the hosts
of 19 compact radio sources at $z\leq0.85$, where the covering factor (Equation
\ref{e1}) is high.} found toward what are generally referred to as
``red quasars''. In these, the reddening of the quasar light is
believed to be due to the presence of dust along the line of sight to
the quasar (e.g. \cite{wfp+95}). These are therefore more likely to
have high molecular abundances (e.g. \cite{wc96a}) but, due to the
dust obscuration, are very difficult objects in which to obtain high
resolution optical spectra: Examples of these are the 4 known
molecular absorbers (\S \ref{currentmol}) of which PKS~1413+135 \&
B3~1504+377 are associated with the host and B0218+357 \&
PKS~1830--210 are gravitational lenses: All have background AGNs with
colours of $V - K >5$, while typical bright quasars (including those
with intervening DLAs) have far bluer colours, $V - K < 3.4$. In using
red quasars, we should bear in mind, however, that:
\begin{itemize}
	\item Host absorption systems sometimes have strong velocity fields,
due to their proximity to the AGN, which may result in various
lines arising in regions of differing velocities. It is thus important
to obtain multiple estimates in individual sources.
	\item The structure of the background emitting region can be
different between centimetre and millimetre wavelengths; this can be
important for gravitational lenses since the light paths of the lines
might be quite different. It is thus more prudent to compare lines arising at
similar frequencies, such as the 18 cm OH ground state lines.
\end{itemize}

\subsection{Atomic lines}
\label{21cm}

Although millimetre lines are absent in DLAs (see \S \ref{mol}), and
the relative incidence of 21 cm absorption is considerably lower than
in red quasars ($\approx10\%$, cf.  $\approx80\%$ \cite{cmr+98}), DLAs
and sub-DLAs constitute a large ($>300$) sample\footnote{Comprising of
150 confirmed DLAs (e.g. \cite{pgw03a}) and $>150$ sub-DLAs/candidate
systems.} of high column density ($N_{{\rm HI}}\gapp10^{20}$~\scm)
absorbers at known redshifts. Of these, although only 31 have been
thoroughly searched (\S \ref{targets}), there are 58 known DLAs and a
further 29 sub-DLAs which occult radio-loud quasars, where we define
this as $\geq0.5$ Jy, the minimum flux density illuminating a DLA
currently detected in 21 cm absorption (see \cite{cmp+03}).

For an optically thin, homogeneous
cloud, the column density (\scm) of the absorbing gas is given by
\begin{equation}
N_{\rm HI}=1.823\times10^{18}.\frac{T_{\rm spin}}{f}\left.\int\right.\tau dv
\label{e1}
\end{equation}
where $T_{\rm spin}$ is the spin temperature of the gas (K), $f$ is
the covering factor of the continuum flux and $\int\tau dv$ is the
integrated optical depth of the line (\kms). Applying the $3\sigma$
opacities of $\tau\approx1 - 20 \times10^{-4}$ obtained after one hour
of integration (\S \ref{tech}) with the SKA, gives $N_{{\rm
HI}}\approx2 - 20 \times10^{14}.\frac{T_{\rm spin}}{f}$ \scm ~per unit
\kms ~ line-width\footnote{Note that for a single cloud in
thermodynamic equilibrium, the spin temperature is approximately equal
to the kinetic temperature of the gas, giving $T_{\rm
s}\approx22\times{\rm FWHM}^2$. That is, for $T_{\rm spin}\geq100$ K
the line-width (FWHM) is expected to be $\geq2$
\kms.}$^{,}$~\footnote{Since for continuum fluxes of $S>0.05$ Jy, the
antenna temperature exceeds the receiver temperature of the SKA, this
estimate will hold for all values significantly greater than this as
$T_{\rm sys}=T_{\rm ant}+T_{\rm rec}\propto S$ and
$\tau\approx\frac{\sigma}{S}$ ($\tau<<1$).} over $z_{\rm abs} = 0 -
6$. For $T_{\rm spin}\approx100$ K, a $\sim10$ \kms ~wide\footnote{In
the case of \HI ~these can be $\approx4 - 50$ \kms ~in DLAs
(\cite{wbcj78,lb01}) and $\approx15 - 100$ \kms ~in
red quasars (e.g. \cite{cmr+98}).} 21 cm absorption line, at 1 \kms
~resolution, would be detectable to $N_{{\rm HI}}\gapp10^{17}$ \scm
~after one hour. That is, for the first time in Lyman-limit systems,
thus increasing the number of potential 21 cm absorbers for which the
redshifts are currently available by 3 orders of magnitude. For DLAs
(where $N_{\rm HI}\geq2\times10^{20}$~cm$^{-2}$) we obtain the limit
$\frac{T_{\rm spin}}{f}\gapp10^4$ K ($\sim10$ \kms ~line-width), thus
covering a large range of spin temperatures and covering factors in
the known systems which occult radio-loud quasars\footnote{The
redshifted frequencies of many of these make them inaccessible to
current telescopes.}. Immediately, this increases the number of
possible 21 cm absorbing DLAs and sub-DLAs by a factor of 5 in the
systems currently known.  Furthermore, these limits significantly
improve the potential of finding absorption in those DLAs much more
weakly illuminated at radio frequencies.

With the high quality spectra obtained from the high column density
absorbers, Voigt profile fitting to the line shape can then be used to
measure the line strength to a fraction of the channel width,
suggesting that even such ``detection'' experiments could obtain a
redshift precision of $\Delta z \approx 3 \times 10^{-6}$. Clearly, in
reasonable integration times, the SKA should be able to measure the
21~cm line redshift to a precision of better than $\sim 10^{-6}$ for
all $\approx 50$ DLAs of the current $z \gapp 2$ sample. 

Similar accuracies of $10^{-6}$ are also possible from measurements of
the redshifts of optical resonance lines in these DLAs.  However,
systematic offsets between the velocities of the optical and 21~cm
lines are likely to dominate the error estimates on individual
systems, which could be of order 5~--~10~\kms, due to small scale
motions in the ISM.  Assuming a dispersion of $\approx 3$~\kms~
between the optical and 21~cm redshifts, implies an additional
redshift error of $\sim 10^{-5}$, far larger than the measurement
error. However, even with only 21~cm detections in the 50~DLAs of the
current $z\gapp2$ sample, a large increase in 
precision is obtained: For an assumed median redshift of 2.5, the $1\sigma$
sensitivity to changes in $\Delta x/x = 4 \times 10^{-7}$. Note that
the quadratic dependence of $x$ on $\alpha$ implies that the
sensitivity to a variation in $\alpha$ is already comparable to the
sensitivity of the Oklo measurement (\S \ref{intro}).

\subsection{Molecular lines}
\label{mol}

As mentioned previously (\S \ref{targets}), by using optically
selected objects, such as DLAs, we are selecting against dusty
(i.e. visually obscured) objects which are more likely to harbour
molecules in abundance (e.g. \cite{fp93,lps03}). In fact, the DLAs in
which \MOLH~ has been detected ($\approx10\%$ of the sample), have
molecular hydrogen fractions, ${\cal F}\equiv 2N({\rm H}_2)/[2N({\rm
H}_2)+N($H{\sc \,i}$)]$, in the range $-6 < \log_{10}{\cal F} < -2$
\cite{lps03}. These optical searches are generally limited to above
the atmospheric cut-off of the Lyman and Werner \MOLH-bands in the UV
($z_{\rm abs}\gapp1.8$).  However, although not as sensitive as in the
optical domain, searches for absorption in the rotational transitions
of CO and HCO$^+$ show that molecular fractions of $\log_{10}{\cal F}
\lapp -1$ are expected at low redshift (see \cite{cmpw03} and
references therein). These low molecular fractions are not surprising
given that these samples are drawn from flux-limited optical
surveys. Surveys for 21~cm absorption with the SKA will, however, give
rise to samples of ``DLAs'' unbiased by dust extinction; some of these
are likely to have reasonable molecular fractions, yielding many
systems with detections of both \HI~and molecular lines.

\subsubsection{Millimetre transitions}
\label{mmdla}

With its frequency cut-off of 25 GHz, the SKA will be able to observe
the strongest absorbing molecules detected in the 4 known systems,
HCN, HCO$^+$ and CO, at $z_{\rm abs}\gapp2.6$. At the low end of this
range, a one hour integration at 1 \kms ~ resolution
gives a $3\sigma$ opacity of $\tau\approx3\times10^{-4}$ (\S
\ref{tech}) for background continuum flux of 0.1~Jy. For rotational
lines the total column density of each molecule is given by
\begin{equation}
 N_{\rm mol}=\frac{8\pi}{c^3}\frac{\nu^{3}}{g_{J+1}A_{J+1\rightarrow J}}\frac{Qe^{E_J/kT_{\rm x}}}{1-e^{-h\nu/kT_{\rm x}}}
\left.\int\right.\tau dv
\label{e2}
\end{equation}
where $\nu$ is the rest frequency of the $J\rightarrow J+1$
transition, $g_{J+1}$ and $A_{J+1\rightarrow J}$ are the statistical
weight and the Einstein A-coefficient of the transition, respectively,
$Q = \sum^{\infty}_{J=0}g_{J}~e^{-E_J/kT_{\rm x}}$ is the partition
function, for the excitation temperature, $T_{\rm x}$, and $\int\tau
dv$ is the upper limit of the velocity integrated optical depth of the
line.  Assuming a typical ``dark cloud'' excitation temperature of
$T_{\rm x}\gapp20$ K (i.e. 10 K at $z=0$), the SKA will give column
density sensitivities of $N_{{\rm CO}}\approx6\times10^{12}$ \scm,
$N_{{\rm HCO+}}\approx5\times10^{9}$ \scm ~and $N_{{\rm
HCN}}\approx10^{10}$ \scm ~per unit line-width\footnote{The 4 known
redshifted millimetre absorbers (\S \ref{currentmol}) have line-widths
$\lapp20$ \kms.} after one hour of integration.

In terms of sensitivity, these limits are comparable to the current
optical (CO electronic) searches in DLAs (see
\cite{cmpw03})\footnote{Note that, using the UV Werner band, optical
CO searches are restricted to $z_{\rm abs}\gapp1.6$, cf. $z_{\rm
abs}\gapp3.6$ with the SKA.}, and due to the steep molecular fraction
evolution in these objects \cite{cwmc03}, the detection of millimetre
transitions is expected to remain beyond our grasp. However, for
$N_{{\rm HI}}\gapp10^{20}$ \scm ~and $N_{{\rm H}_2}\sim 10^4 N_{{\rm
CO}}$ (again for dark clouds, e.g. \cite{wc95}), we obtain
sensitivities to molecular fractions of ${\cal F}\gapp10^{-4}$ (or
${\cal F}\gapp10^{-2}$ using the, perhaps more reliable \cite{ll00},
conversion of $N_{{\rm HCO^+}}=2-3\times10^{-9} N_{{\rm H}_{2}}$),
cf. ${\cal F}=0.3 - 1.0$ for the 4 known millimetre systems
(e.g. \cite{cw98b}). Note that since these estimates are per unit
line-width the sensitivities will decrease directly as the FWHM of the
line\footnote{For example, in a detection experiment, where a
resolution (and FWHM) of 10 \kms ~may be used, these sensitivities
would decrease by a factor of $\leq\sqrt{10}$.}.

One foreseeable problem is the possibility of any cosmological
evolution in the molecular hydrogen fraction: According to
\cite{cwmc03}, \MOLH-bearing DLAs exhibit a steep decrease in relative
molecular hydrogen abundance with redshift. If such a trend generally
exists, due to the restriction of searching for $z_{\rm abs}\gapp2.6$
millimetre absorption, the full sensitivity of ${\cal F}\lapp10^{-4}$ may be
required\footnote{Also, for a given
velocity resolution the sensitivity decreases as the square root of
the observed frequencies (\S \ref{tech}) further compounding the difficulty of
detecting molecules at high redshift.}. This problem may be
circumnavigated by searching for centimetre lines at $z_{\rm
abs}\lapp2.6$, which we discuss next.

\subsubsection{OH lines}
\label{ohdla}

Although restrictive in its redshift coverage of millimetre lines, at
centimetre wavelengths many times as many sources at a given flux
density are expected than at higher frequencies. Additionally, 
since the instantaneous bandwidth is a quarter of the band centre frequency,
complete searches for 18 cm OH to $z \lapp 16$ can be performed
in relatively few scans. This combined with the large
field of view at low frequencies, makes the SKA the ideal instrument
with which to perform blind searches for OH absorption.

For an optically thin cloud in thermal equilibrium, the OH column density 
is related to the excitation temperature and 1667~MHz optical 
depth by (e.g. \cite{pkn+04})
\begin{equation}
N_{\rm OH}\approx2.38\times10^{14}.\frac{T_{\rm x}}{f}\left.\int \right.\tau dv
\label{eqn:oh}
\end{equation}
which, again for a flux density of 0.5 Jy, gives a $3\sigma$ sensitivity of
$N_{\rm OH}\approx3 - 30\times10^{10}\frac{T_{\rm x}}{f}$ \scm ~over
$z_{\rm abs} = 0 - 7$, per \kms ~line-width. Since the line-widths can
range anywhere from $\approx10 - 200$ \kms ~in the 4 known systems (\S
\ref{currentmol}), pessimistically {\it very high resolution} spectra
of very wide lines will be detected by the SKA to $N_{\rm
OH}\sim10^{13}\frac{T_{\rm x}}{f}$ \scm, cf.  $N_{{\rm
OH}}\sim1-12\times10^{14}\frac{T_{\rm x}}{f}$, currently detected at
{\it low resolution} in the known systems ($z_{\rm abs} \leq0.9$).

For $N_{\rm OH}\approx 30N_{\rm HCO^+}$ (\S \ref{CO}) and using the
$N_{{\rm HCO+}}$--$N_{{\rm H}_{2}}$ conversion ratio quoted above,
gives a sensitivity to ${\cal F}\gapp10^{-2}$ after one hour, in the
low redshift regime. Since redshifts of $z_{\rm abs}\lapp2.6$ can be
observed in OH 18 cm, any molecular fraction evolution is of
considerably less consequence than in the millimetre case.

\section{SUMMARY}

To summarise, with its high sensitivity, wide field of view and broad
\begin{table*}
\caption{Summary of the various combinations of fundamental constants
which can be constrained from various spectral lines, where $g_p$ is the  proton g-factor, $\alpha\equiv e^2/\hbar c$ is the fine structure constant and  $\mu\equiv m_e/m_p$ the ratio of
electron/proton masses.}
\label{t1}
\newcommand{\m}{\hphantom{$-$}}
\newcommand{\cc}[1]{\multicolumn{1}{c}{#1}}
\renewcommand{\tabcolsep}{2pc} 
\renewcommand{\arraystretch}{1.2} 
\begin{tabular}{@{}l l c}
\hline
Transition &  ``Anchor''  & Constrained quantity\\
\hline
\HI~21cm    & Metal-ion (optical)     & $ g_p \mu \alpha^2$ \\
	    & HCO$^+$     & $g_p \alpha^2$      \\
            & OH 18cm ($\nu_{1665} + \nu_{1667}$) & $ g_p [\alpha^2/\mu]^{1.57} $\\
	    & & \\
OH 18cm ($\nu_{1665} + \nu_{1667}$)	 & HCO$^+$     & $\mu^{1.57} \alpha^{-1.14}$ \\
    & OH 18cm ($\nu_{1665} - \nu_{1667}$) & $ g_p [\alpha^2/\mu]^{0.13}$\\
 	    & OH 18cm ($\nu_{1720}-\nu_{1612}$) & $g_p [\alpha^2/\mu]^{1.85}$ \\
	    & OH 6cm   & $[\alpha^2/\mu]^{-2.06} $\\
\hline
\end{tabular}
\end{table*}
frequency coverage (unaffected by RFI), the SKA will revolutionize
radio studies of the cosmological evolution of various fundamental
constants (see Table \ref{t1}).
The SKA will be unparalleled in detecting these lines in:
\begin{itemize}
  \item Optically selected QSO absorbers: Being of low dust content
these sources are best targeted for 21 cm absorption. In this
transition, the SKA truly excels and we expect detections in systems
with column densities as low as $N_{{\rm HI}}\sim10^{17}$ \scm
~(i.e. Lyman limit systems) or in DLAs with spin temperature/covering
factor ratios as high as $ \frac{T_{\rm spin}}{f}\sim10^4$. Even with
the currently known redshifted optical absorption systems, this will
increase the number of known 21 cm absorbers many-fold.
       \item Red quasars: Being visually obscured these objects
	 provide the best targets for molecular absorption searches,
	 but are, by definition, currently much rarer in current
	 searches than optical absorbers.  However, with its extremely
	 wide band-width and field of view, the SKA has the potential
	 to discover many more systems invisible to optical surveys,
	 via the \HI ~21 cm and OH 18 cm lines.
	 
        Currently there are only 4 redshifted radio molecular
	absorbers known.  The discovery of more of these systems will
	prove invaluable in studying the physics and chemistry of the
	early Universe. Furthermore, molecular absorption lines are
	extremely important in constraining fundamental constants since:
	 \begin{itemize}
	 \item The spectral resolution available in radio spectroscopy
	   is generally much better than the optical, thus giving a
	   measurement better matched for comparison with the \HI
	   ~line. This will allow estimates of $\Delta x/x \sim few
	   \times 10^{-7}$ ($x\equiv g_p \mu \alpha^2$).
	   \item These transitions allow the contribution of the
	     proton g-factor to be separated out and give more
	     combinations with which to determine the relative
	     contribution of each constant to any measured
	     change (Table \ref{t1}). Although the frequency coverage will exclude the
	     detection of millimetre absorption below redshifts of
	     $z_{\rm abs}\approx2.6$, it will excel in finding OH
	     which will provide a diagnostic with which to find
	     HCO$^+$ with higher frequency telescopes such as ALMA.
	 \end{itemize}
\end{itemize}
It is therefore seen that these much improved statistics applied to
the various combinations of fundamental constants available will
permit measurements of the cosmological evolution of each the fine
structure constant, the ratio of electron--proton mass and the proton
g-factor. Through measurements of these combinations of constants in
the early Universe, the SKA could provide the means of experimentally
testing current Grand Unified Theories.



\end{document}